\documentclass[twocolumn,showpacs,pre,nofootinbib]{revtex4}
\newcommand{\figurewidth}{\columnwidth}
\newcommand{\beq}{\begin{eqnarray}}
\newcommand{\eeq}{\end{eqnarray}}
\newcommand{\Tr}{{\text Tr}}
\newcommand{\e}{{\text e}}

\usepackage{graphicx,amsmath}

\begin{document}

\title{Exponential Complexity of the Quantum Adiabatic Algorithm \\ for certain
Satisfiability Problems}
\author{Itay Hen}
\author{A.~P.~Young}
%\email{peter@physics.ucsc.edu}
\affiliation{Department of Physics, University of California, Santa Cruz, California 95064}

\date{\today}

\begin{abstract}
We determine the complexity of several constraint satisfaction problems using
the quantum adiabatic algorithm in its simplest implementation. We do so by
studying the size dependence of the gap to the first excited state of
``typical'' instances. We find that at large sizes $N$, the complexity
increases exponentially for all models that we study. We also compare our results
against the complexity of the analogous classical algorithm WalkSAT and show
that the harder the problem is for the classical algorithm the harder it is
also for the quantum adiabatic algorithm. 
\end{abstract}

\maketitle

\section{Introduction}
\label{sec:intro}

Theoretical research on quantum computing is motivated by the exciting
possibility that quantum computers are inherently more efficient than
classical computers due to the advantages that the laws of quantum mechanics
provide, such as superposition, interference and entanglement. Besides the
great effort of research towards the physical realization of these devices, a
lot of activity has been devoted to the development of algorithms that could
use quantum properties to achieve better efficiency in performing
computational tasks with respect to classical devices.

Perhaps the best example to date for the superiority of quantum computers over
classical ones is given by Shor's algorithm \cite{shor:94} for integer
factorization, which solves the problem in polynomial time, whereas the best
classical algorithm takes a time which is exponential in (a fractional power
of) the problem size.

A rather general approach to solve a \textit{broad range} of hard optimization
problems using a quantum computer has been proposed by Farhi {\it et al.}
\cite{farhi_long:01}.  Within the framework of this new approach, which was
given the name the Quantum Adiabatic Algorithm (QAA), the solution to an
optimization problem is encoded in the ground state of a Hamiltonian
$\hat{H}_p$. 
To find the solution, the QAA prescribes the following course of action. As a first
step, the system is prepared in the ground state of another Hamiltonian
$\hat{H}_d$, commonly referred to as the driver Hamiltonian.  The driver
Hamiltonian is chosen such that it does not commute with the problem
Hamiltonian and has a ground state which is fairly easy to prepare. 
As a next step,
the Hamiltonian of the system is slowly modified from $\hat{H}_d$ to
$\hat{H}_p$, using the linear interpolation, i.e.
\begin{equation}
\hat{H}(s)=s \hat{H}_p +(1-s) \hat{H}_d \,,
\end{equation}
where $s(t)$ is a parameter varying smoothly with time,
from $0$ at $t=0$ to $1$ at the end of the algorithm,
$t=\mathcal{T}$.  If this process is done slowly enough, the
adiabatic theorem of Quantum
Mechanics (see, e.g., Refs.~\cite{kato:51} and~\cite{messiah:62})
ensures that the system will stay close to the ground state of the
instantaneous Hamiltonian throughout the evolution, so that one finally
obtains a state close to the ground state of $\hat{H}_p$.  At this point,
measuring the state will give the solution of the original problem with high
probability. 

The running time $\mathcal{T}$ of the algorithm determines the
efficiency, or complexity, of the QAA.  An upper bound for the complexity can
be given in terms of the eigenstates $\{ | n
\rangle \}$ and eigenvalues $\{E_n \}$ of the Hamiltonian,
by \cite{wannier:65,farhi:02}
\begin{equation}
\mathcal{T} \gg \hbar \, {| \textrm{max}_{s} V_{10}(s)| \over
(\Delta E_{\textrm{min}})^2} \,,
\end{equation} 
where $\Delta E_{\textrm{min}}$ is the minimum of the first
excitation gap
$\Delta E_{\textrm{min}} = \textrm{min}_{s} \Delta E$
with $\Delta E = E_1-E_0$,
and $V_{n 0} = \langle 0 | {\textrm d} \hat{H} / {\textrm d} s | n\rangle$.

Typically, matrix elements of $\hat{H}$ scale as a low polynomial of the
system size $N$, and the question of whether the complexity depends
polynomially or exponentially with $N$ 
therefore depends on how the minimum gap $\Delta
E_{\textrm{min}} $ scales with $N$.  This means that if the gap becomes
exponentially small at any point in the evolution, then the computation
requires an exponential amount of time, rendering the QAA inefficient.  The
dependence of the minimum gap on the system size for a given problem is
therefore a central issue in determining the complexity of the QAA. 

The most interesting unknown about the QAA to date is thus whether or not it
could solve in polynomial time ``hard'' sets of problems -- those which belong
to the NP-complete category  \cite{garey:97} and for which all known
algorithms take an exponential amount of time (exponential complexity) at
least in the worst case. 
While early studies of the QAA done on very small systems ($N \leq 24$)
\cite{farhi_long:01,hogg:03} provided some preliminary numerical evidence 
that the time required to solve one such NP-complete problem does scale only
polynomially, roughly as $N^2$, several later studies gave evidence
that this may not be the case.

Refs.~\cite{farhi:02,farhi:08} show that
adiabatic algorithms can fail if one does not choose the initial Hamiltonian
carefully by taking into account the structure of the problem. Altshuler {\it
et al.}~\cite{altshuler:09} also argued that adiabatic quantum optimization
will fail in general for random instances of NP-complete problems.
However, the arguments of Altshuler {\it et al.}~have been criticized by
Knysh and Smelyanskiy \cite{knysh:11}.

In addition, Young {\it et al.}~\cite{young:08,young:10} recently examined the
1-in-3 ``constraint satisfaction'', or SAT, problem (to be explained 
in the next section) and showed that very small gaps could appear in
the spectrum of the Hamiltonian due to an avoided crossing between the ground
state and another level corresponding to a local minimum of the optimization
problem.  This `bottleneck' was shown to appear in a larger and larger
fraction of the instances as the problem size $N$ increases, indicating the
existence of a first order quantum phase transition. This leads to an
exponentially small gap of a \textit{typical} instance,
and therefore also to the failure of adiabatic
quantum optimization. Other studies that considered this model have
found an exponential complexity \cite{znidaric:05,znidaric:06} for
\textit{particularly hard} instances of small size.

It is not yet clear however to what extent the above behavior found for
1-in-3 SAT is general and whether it is a feature inherent to the QAA that will
plague most if not all problems fed into the algorithm or something far more
restricted than this. 
Previous work~\cite{jorg:08,jorg:09,jorg:10b} had argued that
a first order quantum phase transition occurs for a broad
class of random optimization models.
To gain further insight into this matter we study here three optimization
problems which had previously been suggested
~\cite{zdeborova:08a,zdeborova:08b,jorg:09} as good potential candidates for
detailed investigation.

The problems we study are of the ``constraint satisfaction'' type.
For these, one asks a questions for which there is a ``yes'' or ``no'' answer,
namely
whether there is an assignment of $N$ bits which satisfies all of $M$ logical
conditions (clauses). An energy is assigned to each clause such that
it is zero if the clause is satisfied and positive if it is not. 

The first two problems we focus on in this paper are ``locked'' problems -- a
term first introduced by Zdeborov\'a and
M\'ezard~\cite{zdeborova:08a,zdeborova:08b} for problems with instances having
the following two properties: (i) every variable is in at least two clauses,
and (ii) one can not get from one satisfying assignment to another by flipping
a single bit. In fact, it was argued that typically order $\ln N$ bits needs
to be flipped to go from one solution to another.  These locked problems have
several properties that make them
eminently suitable as benchmarks.  They  are analytically ``simple'' (or at
least simpler than previously studied models such as random K-SAT), but are
computationally hard. Also, fluctuations between instances are smaller than
with ``unlocked'' problems. 
Specifically, we study here the complexity of the QAA for the locked 1-in-3 SAT
and locked 2-in-4 SAT models which belong to the NP-complete category.

In addition, we also compare our results with those of a third model, 
3-regular 3-XORSAT, already considered by J\"org {\it et al.}~\cite{jorg:09},
and Farhi {\it  et al.}~\cite{farhi:11}.
As we shall see, this model, while belonging to the P complexity class  (i.e., it could be solved in polynomial time)
is very hard to solve computationally by general purpose algorithms.

We study these models by analyzing the size dependence of the typical gap by
means of quantum Monte Carlo (QMC) simulations.  The plan of this paper is as
follows: Section~\ref{sec:models} describes the three models that will be
studied. In Sec.~\ref{sec:method} we discuss the manner in which we obtain our
results. These results are presented in Sec.~\ref{sec:results} and our conclusions are
summarized in Sec.~\ref{sec:conclusions}.

\section{Models}
\label{sec:models}

We consider problems of the ``constraint satisfaction'' type, in which
there are $N$ bits (or equivalently, Ising spins) and $M$ ``clauses'' where each
clause is a logical condition on a small number of randomly chosen bits.  A
configuration of the bits (spins) is a ``satisfying assignment'' if it
satisfies all the clauses. 

In encoding this type of problem as a quantum Hamiltonian, each bit variable
is represented in the Hamiltonian by the $z$-component of a Pauli
matrix, $\sigma_i^z$, where $i$ labels the spin.  Each clause is thus
converted to an energy function which depends on the spins associated with the
clause, such that the energy is zero if the clause is satisfied and is
positive (in our case, one) if it is not.  The general structure of the
problem Hamiltonian $\hat{H}_p$ is therefore
\begin{equation}
\hat{H}_p = \sum_{a=1}^{M} \hat{H}_a \,,
\end{equation}
where $a$ is the clause index and $\hat{H}_a$ is the energy associated with
the clause and involves the spins belonging to it. 

Clearly, it is easy to satisfy all clauses if the ratio $\alpha \equiv M/N$ is
small enough. In fact, one expects an exponentially large number of satisfying
assignments in this region.  Conversely, if $M/N$ is very large, with high
probability there will be a conflict between different clauses.  Hence there
is a ``satisfiability transition'' at some value $\alpha_s$ where the number
of satisfying assignments goes to zero.  It is
particularly hard to solve satisfiability problems close to the
transition~\cite{kirkpatrick:94}, so we will work in this region.

Furthermore, when studying the efficiency of the QAA
numerically~\cite{farhi_long:01,young:08,young:10}, it is convenient to
consider instances with a unique satisfying assignment (USA), which, of course,
forces the system to be close to the transition.  
Considering instances with a USA is particularly advantageous for locked problems. 
While for unlocked problems the entropy of solutions at the satisfiability threshold is positive \cite{monasson:96},
for locked problems it approaches zero continuously \cite{zdeborova:08a,zdeborova:08b}. 
This means that while solutions with USA are rare for unlocked problems, they are expected to be among the `typical' instances
for locked problems and therefore locked problems have the
advantage that instances with a USA should be a good representation of
\textit{randomly chosen} instances. Indeed, this is supported by a recent numerical  study \cite{guidetti:11} that found that the
probability of a USA only decreases slowly with $N$ and appears to tend to a nonzero value  as $N \to \infty$. 

We now discuss the different models that will be investigated in this paper.

\subsection{Locked 1-in-3 SAT}
\label{sec:locked_1-in-3}

In the 1-in-3 SAT problem each clause consists of three bits chosen randomly,
and the clause is satisfied if one of the bits is one and the others are zero.
Here we fix the ratio $M/N$ to be the critical value for the satisfiability
transition. According to Table I of Ref.~\cite{zdeborova:08b}, this is equal
to $\alpha_s = 0.789$. Since $M$ has to be an integer we take $M$ to be the
nearest integer to $\alpha_s N$, see Table~\ref{tab:params}.
Note that, if the sites are chosen at random
to form the clauses, the distribution of the degree of the sites (i.e., the
number of clauses involving a site) would be Poissonian. However, locked
instances have a minimum degree of two, so instead we use a truncated
Poissonian distribution~\cite{zdeborova:08b} which is Poissonian except that
the probabilities for zero and one are set to zero.

We study instances with a unique satisfying assignment (USA). For these
instances the gap to the first excited state is of order unity at $s=1$ (and
also of order unity at $s = 0$) so the gap has a
minimum whose value is related to the complexity. For instances with many
satisfying assignments the ground state of the problem Hamiltonian
is degenerate and so the gap to the
first excited state decreases to zero as $s \to 1$. Hence this gap would give no
information about the computer time needed to determine whether there is a state
with zero energy. 

The energy of a clause for the locked 1-in-3 problem is given by:
\begin{eqnarray}
\hat{H}_{a} &=& \frac1{8}\Big(5-\sigma_{a_1}^z - \sigma_{a_2}^z - \sigma_{a_3}^z  \\\nonumber 
&+& \sigma_{a_1}^z \sigma_{a_2}^z +\sigma_{a_2}^z \sigma_{a_3}^z  +\sigma_{a_3}^z \sigma_{a_1}^z  
+3 \sigma_{a_1}^z \sigma_{a_2}^z \sigma_{a_3}^z \Big) \,,
\end{eqnarray}
where $a$ denotes the index of the clause and the $a_i$ ($i=1,2,3$) 
label the participating spins. With this Hamiltonian, the energy
is zero if the clause is satisfied
and is one otherwise.

\begin{table}
\begin{center}
\begin{tabular}{||c|c|c||}
\hline\hline
  & locked 1-in-3 & locked 2-in-4 \\
\hline\hline
N  & M &  M  \\
\hline
16 &  13 &  11 \\
24 &  -- &  17 \\
32 &  25 &  23 \\
40 &  -- &  28 \\
48 &  38 &  34 \\
64 &  51 &  -- \\
96 &  76 &  -- \\
\hline\hline
\end{tabular}
\end{center}
\caption{
Values of $M$ and $N$ for the locked instances.}
\vspace{-0.5cm}
\label{tab:params}
\end{table}

\subsection{Locked 2-in-4 SAT}
\label{sec:locked_2-in-4}

We also consider locked 2-in-4 instances, in which a clause has four bits, and
is satisfied if two are zero and two are one. Unlike the locked 1-in-3 SAT
model discussed above, this model has a symmetry under flipping all the bits.

We fix the ratio $M/N$ to be the critical value for the satisfiability
transition. According to Table I of Ref.~\cite{zdeborova:08b}, this is equal
to $\alpha_s = 0.707$. Again, since $M$ has to be an integer we take $M$ to be
the nearest integer to $\alpha_s N$, see Table~\ref{tab:params}. 

In this problem, the energy of a clause is given by:
\begin{eqnarray}
\hat{H}_{a} &=& \frac1{8}\Big(5
+\sigma_{a_1}^z \sigma_{a_2}^z +\sigma_{a_1}^z \sigma_{a_3}^z +
\sigma_{a_1}^z \sigma_{a_4}^z +\sigma_{a_2}^z \sigma_{a_3}^z
\nonumber\\
& +&\sigma_{a_2}^z \sigma_{a_4}^z +\sigma_{a_3}^z \sigma_{a_4}^z 
- 3 \sigma_{a_1}^z \sigma_{a_2}^z \sigma_{a_3}^z \sigma_{a_4}^z \Big) \,,
\end{eqnarray}
where, as before, $a$ denotes the index of the clause and the $a_i$ ($i=1,2,3,4$)
label the participating spins. For this energy term, a satisfied clause has
zero energy and an unsatisfied one has energy one.

Because of bit-flip symmetry the energy of a state of the problem
Hamiltonian is the same as that of the state obtained by flipping all the
bits. Hence, when we refer to an instance with a ``unique'' satisfying
assignment (USA) for the locked 2-in-4 problem, we will ignore states related
by symmetry (so the true ground state degeneracy is actually two, not one).

\subsection{3-regular 3-XORSAT}
\label{sec:XORSAT}

Another problem we discuss here is the 3-regular 3-XORSAT problem, already
considered by J\"org {\it et al.}~\cite{jorg:09} and
Farhi {\em et al.}~\cite{farhi:11}. In the 3-XORSAT problem,
three bits are chosen to form a clause and the clause is satisfied if their
{\em sum} (mod 2) is a specified value (either 0 or 1). Alternatively, in terms of
spins, the clause is satisfied if the {\em product} of the three
$\sigma_i^z$'s is a specified value (either $-1$ or 1).

We will consider here
the ``3-regular'' case where every bit is in exactly three clauses, a model
which turns out to be \textit{precisely} at the satisfiability threshold.
Note that this implies $M = N$.
Again, the problem to be solved is whether there is an assignment of the 
bits which satisfies all the clauses. Interestingly, since this problem just
involves linear algebra (mod 2), the satisfiability problem can be solved in
polynomial time using, for example, Gaussian elimination. However, as is already well known (see, e.g., \cite{franz:01,ricci-tersenghi:11}) 
and will also become evident soon, the problem is very hard for general purpose algorithms.
Furthermore, if there is
\textit{no} satisfying assignment, no known polynomial time algorithm
will determine the minimal number of unsatisfied clauses, a problem 
known as MAX-XORSAT.

As usual, we consider instances with a USA. Fortunately, these are \textit{a
nonzero fraction}, about 0.285~\cite{jorg:09}, of the total, so the USA
instances should be a good representation of randomly chosen ones. 
For XORSAT instances with a USA, it is not difficult
to show that one can gauge transform any instance into one in which the sum of
the bits of every clause is equal to 0 (mod 2). The USA is then all bits equal
to 0, (a ``ferromagnetic" ground state in statistical physics language).
Although this ground state is ``trivial", we shall see that it is very hard to
find using general purpose algorithms including the QAA. 

The energy of a clause in this model is:
\begin{equation}
\hat{H}_a = \frac1{2}\left(1- 
\sigma_{a_1}^z \sigma_{a_2}^z \sigma_{a_3}^z\right) \,,
\end{equation}
where again $a$ denotes the index of the clause and the $a_i$ ($i=1,2,3$) 
label the participating spins.

%The sizes we study are $M = N = 12, 16, 20, 24, 32$ and $40$.
 
\subsection{The driver Hamiltonian}
Before moving on, we note that the driver Hamiltonian we choose here is
perhaps the simplest possible choice,
\begin{equation}
\hat{H}_d = \frac1{2} \sum_i \left( 1-\sigma_i^x \right) \,,
\end{equation}
where  $\sigma_i^x$ is the $x$-component Pauli matrix acting on spin $i$.
This corresponds to a
transverse field of equal size on all sites. Its ground state is a
uniform superposition of all $2^N$ states of the computational
(i.e.~$\sigma^z$) basis.

\section{\label{sec:method}Method}

As was already mentioned, the complexity of the QAA algorithm is determined by
the size dependence of the ``typical'' minimum gap of the problem. Following
Refs.~\cite{young:08} and \cite{young:10}, we analyze the size-dependence of
these gaps for each of the problems discussed in the previous section by
considering typically 50 instances for each size, and then extracting the
minimum gap for each of them.  As a next step, we take the median value of the
minimum gap among the different instances for a given size to obtain the
``typical'' minimum gap.

\begin{figure}
\begin{center}
\includegraphics[width=\figurewidth]{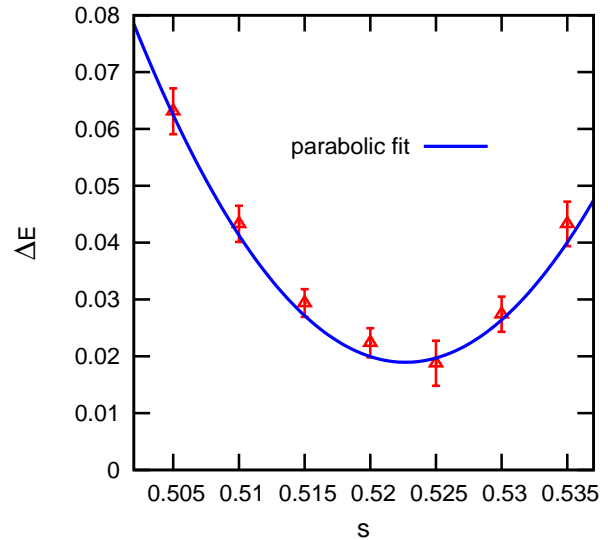}
\caption{(Color online) 
Gap to the first excited state as a function of $s$ for one instance of the
locked 1-in-3 SAT problem.  The line is a quadratic fit to the data points from
which the location, $s=0.523$, and value, $\Delta E_{\textrm min}=0.207$, of the
minimum gap are obtained.  Here, $N=64$ and $\beta=1024$.}
\label{fig:gapS}
\vspace{-0.7cm}
\end{center}
\end{figure}

To find the minimum gap for a specific instance of a specific problem, we
perform
quantum Monte Carlo simulations for a range of $s$
values that bracket the minimum gap. 
For each of the
studied $s$ values we extract the gap and interpolate the minimum value using
a simple quadratic fit.
An illustrative
example of this is given in Fig.~\ref{fig:gapS}

In cases where we find that the mesh of $s$ values is either too crude or does
not bracket the minimum gap, a second round of simulations, with a more
appropriate mesh of $s$ values, is launched.

\subsection{The quantum Monte Carlo technique}

To study the behavior of the typical minimum gap for large ($N >24$) system
sizes for which exact diagonalization routines are unfeasible, we employ a
continuous-time quantum Monte Carlo technique.  As the name indicates,
this technique is based on sampling the $2^N$ states of the Hilbert
space, so there are therefore statistical errors stemming from the nature of
the procedure. However, Monte Carlo methods provide the only numerical method
available for investigating large system sizes.

The specific method we use in this study is known as the stochastic series
expansion (SSE) algorithm \cite{sandvik:99,sandvik:92} which involves a
Taylor series expansion of the partition function $\Tr[\e^{-\beta \hat{H}}]$
and uses a discrete representation of continuous imaginary time.
This discretization however does not introduce errors into the algorithm as is
the case in the alternative path-integral formulation, where one usually
performs a Trotter-type discretization of imaginary time, see e.g.
Refs.~\cite{young:08,young:10}, though formulations in continuous imaginary
time also exist \cite{farhi:09,krzakala:08}.  Here $\beta$ is the inverse
temperature $1/T$ (in our units $k_B=1$). 

The SSE algorithm has several properties that are very useful in addressing
the problems we focus on in this study. Firstly, it works in continuous imaginary time as
discussed above. Secondly, it allows not only local updates of
system configurations but also global cluster updates, which in most cases
prove to be more efficient than single-spin-flip updates. These global updates
are achieved by dividing the configurations of the system produced by the QMC
into clusters and then flipping a fraction of them within each sweep of the
simulation \cite{sandvik:03}.  An important bonus of cluster updates is
the existence of ``improved estimators'' for determining time-dependent
correlation functions, for which the signal to noise is much better than with
conventional measurements.

In addition, we speed up equilibration by implementing
``parallel tempering'' \cite{hukushima:96},
where simulations for different values of $s$ are
run in parallel and spin configurations with adjacent values of $s$ are
swapped with a probability satisfying the detailed balance condition.
Traditionally, parallel tempering is performed for systems at different
temperatures, but here
the parameter $s$ plays the role of (inverse) temperature.

We extract the gap from imaginary time-dependent correlation functions.
However, in the locked 2-in-4 case, where the problem has bit-flip symmetry,
this is tricky using the standard SSE algorithm and, as discussed next,
we will use a different approach.

\begin{figure}
\begin{center}
\includegraphics[width=\figurewidth]{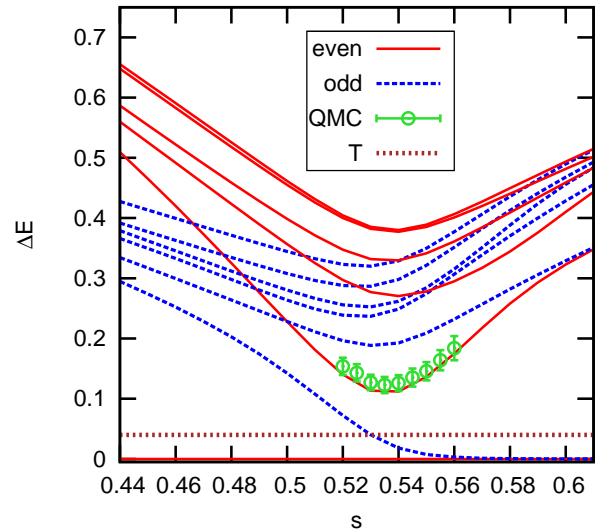}
\caption{(Color online) 
Energy gaps to even (solid, red) and odd (dashed, blue) excited states for an $N=16$
instance of the locked
2-in-4 problem, which has bit-flip symmetry as discussed in the text.
The dotted line shows
a characteristic value of 
another important energy scale in the problem, temperature. In the region where the gap to
the first even state has a minimum, the gap to the first odd state becomes
very small and is inevitably thermally populated. Hence, odd-odd gaps appear
in this region as well as even-even gaps. This is the reason why we use a
non-standard Monte Carlo algorithm for this problem which projects out the
symmetric subspace, so only even-even gaps are present in the data. The figure also
shows the gap obtained from the even-subspace projected QMC
in vicinity of the minimum. It
agrees with exact diagonalization within the error bars.}
\label{fig:even_odd}
\vspace{-0.7cm}
\end{center}
\end{figure}

The difficulty arises for the following
reason.
Eigenstates of
the Hamiltonian are either even or odd under bit-flip symmetry
(in particular, the ground state is
even). In the $s \to 1$ limit, states occur in even-odd pairs with an exponentially small
gap (see Fig.~\ref{fig:even_odd} for an illustration).  Therefore, the
quantity of interest is the gap to the first {\it even} state. 
We consider correlation
functions of even quantities, so there are only matrix elements between states
of the same parity.  However, the lowest odd level becomes very close to the
ground state near where the gap to the first even excited state has a minimum,
see Fig.~\ref{fig:even_odd}. Hence this lowest odd state becomes thermally populated,
with the result that
odd-odd gaps are present in the data as well. 

We have eliminated these undesired
contributions by projecting out the symmetric subspace
of the Hamiltonian. 
A way of doing this projection at zero temperature was
proposed independently by Eddie Farhi \cite{fa} and Anders Sandvik
\cite{anders}. In standard quantum Monte Carlo simulations one imposes {\em
periodic} boundary conditions in imaginary time $\tau$ at
$\tau = 0$ and
$\beta$. To project out the symmetric subspace one imposes, instead, {\em
free} boundary conditions \cite{free_refs} at $\tau = 0$ and $\beta$. The properties of the
symmetric subspace can then be obtained, for $\beta \to \infty$, by
measurements far from the boundaries.
We have incorporated this idea into the SSE scheme, and use this modified
algorithm in the simulations of the locked 2-in-4 problem.

\begin{figure}
\begin{center}
\includegraphics[width=\figurewidth]{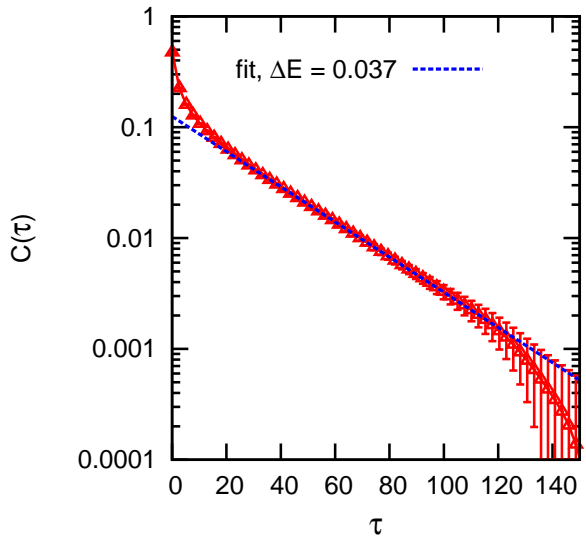}
\caption{(Color online) 
A log-linear plot of a time dependent correlation function for an instance of
the locked 1-in-3 SAT problem with $N = 64$ spins, $\beta=1024$,
near the minimum gap at
$s=0.54$. The energy gap is the negative of the slope at large values of
imaginary-time $\tau$. A fit gives $\Delta E=0.037$.}
\label{fig:ctau}
\vspace{-0.7cm}
\end{center}
\end{figure}

To verify that our implementations of the SSE methods are
accurate, we have compared their results with corresponding
exact diagonalization results on small system sizes. The results agree within
the error bars. A comparison of the gap
is shown in Fig.~\ref{fig:even_odd}. The careful will reader note
that the QMC data is slightly but consistently above the diagonalization results.
This is due to contributions from higher excited states at short times which
increase the value of the time-dependent correlation function used to extract
the gap in this limit, see Fig.~\ref{fig:even_odd} and Sec.~\ref{sec:gap}. The
effect is small even for the $N=16$ data shown in Fig.~\ref{fig:even_odd}, and we
expect it to be smaller still for larger sizes near the minimum gap,
since the gap is smaller so there is a larger region with straight-line
behavior in plots like
Fig.~\ref{fig:ctau}.

\subsection{Extraction of the system gap}
\label{sec:gap}

The gap of the system for a given instance and a given $s$ value is extracted 
by analyzing measurements of (imaginary) time-dependent correlation functions
of the type
\begin{equation}
C_A(\tau) = \langle \hat{A}(\tau) \hat{A}(0) \rangle -\langle A \rangle^2  \,,
\end{equation}
where the operator $\hat{A}$ is some measurable physical quantity.  In
practice, we found it useful to construct superpositions of such
correlation functions. Typically we use linear combinations of
correlation functions of the operators $\sigma_i^z$ or $\sigma_i^z \sigma_j^z$
where $i$ and $j$ run from 1 to $N$ and
label the spins.  The evaluation of $\langle A \rangle^2$
in the above equation is computed from the product $\langle A
\rangle^{(1)} \langle A \rangle^{(2)}$ where the two indices correspond to
different independent simulations of the same system. This eliminates the
bias stemming from straightforward squaring of the expectation value. 

In the low temperature limit, $\Delta E \ll T$ where $\Delta E = E_1 - E_0$,
the system is in its ground state so
the imaginary-time correlation function is given by
\begin{equation}
C_A(\tau) = \sum_{m=1} |\langle 0 | \hat{A} | m \rangle|^2
\left( e^{-\Delta E_m \tau} + e^{-\Delta E_m (\beta - \tau)} \right)  \,,
\end{equation}
where $\Delta E_m = E_m - E_0$. At 
long times, $\tau$, the correlation function is dominated by the
smallest gap, $\Delta E \equiv \Delta E_1$,
(as long as the
matrix element $|\langle 0 | \hat{A} | 1 \rangle|^2$ is nonzero).  On a log-linear plot
$C_A(\tau)$ then has a region where it is a straight line whose slope
is the negative of the gap. This can therefore be easily extracted by simple
linear fitting.

An illustration of the above procedure is depicted in
Fig.~\ref{fig:ctau} showing one of the correlation functions measured and
analyzed for the locked 1-in-3 problem for $N=64$, $\beta=1024$ and $s=0.54$.
The gap in this case is $\Delta E=0.037$. 

\section{\label{sec:results}Results}
\subsection{Results from the QAA}

\begin{figure}
\begin{center}
\includegraphics[width=\figurewidth]{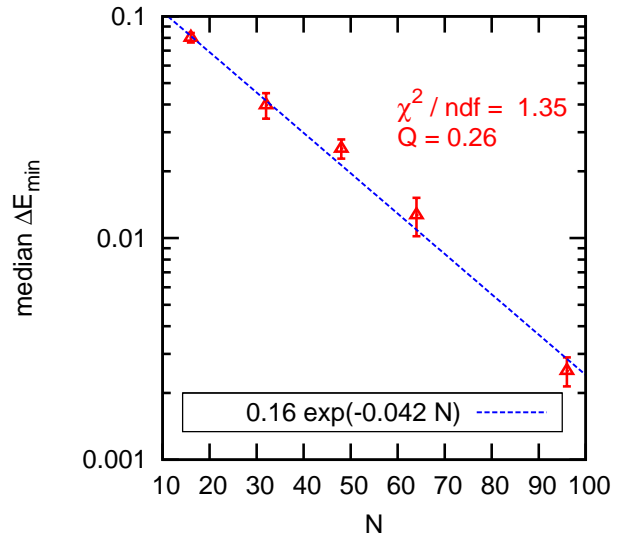}
\caption{(Color online) Median minimum gap on a log-linear scale for the locked 1-in-3
problem. The straight-line fit is good indicating 
the exponential complexity of the QAA for this problem.}
\label{fig:L13DE}
\vspace{-0.7cm}
\end{center}
\end{figure}
\begin{figure}
\begin{center}
\includegraphics[width=\figurewidth]{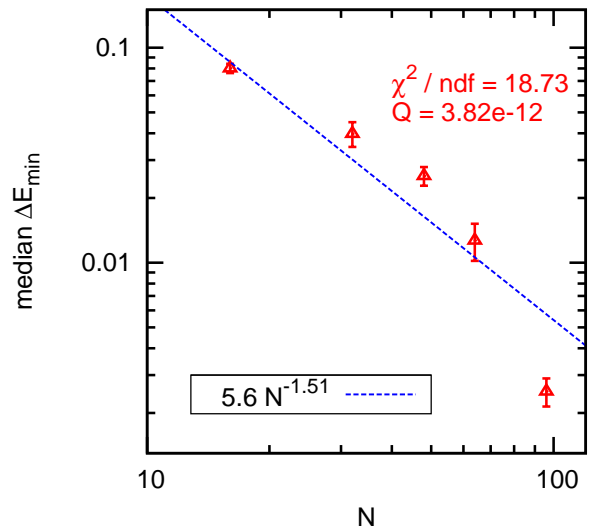}
\caption{(Color online) Median minimum gap on a log-log scale for the locked 1-in-3
problem. The straight-line
fit is extremely poor indicating that the minimum gap for this problem is
not polynomial in the system size.}
\label{fig:L13DE_loglog}
\vspace{-0.7cm}
\end{center}
\end{figure}

We show results for the median minimum gap as a function of size for the
locked 1-in-3 problem in Fig.~\ref{fig:L13DE} (log-lin) and
Fig.~\ref{fig:L13DE_loglog} (log-log). A straight line fit works very well for the
log-lin plot (goodness of fit parameter $Q = 0.26$) but very poorly for the
log-log plot ($Q = 3.8 \times 10^{-12}$). This
provides strong evidence that the minimum gap is exponentially small in 
the system size, and so the complexity of the QAA (at least in the simplest
version considered here) is exponentially large in the system size.

\begin{figure}
\begin{center}
\includegraphics[width=\figurewidth]{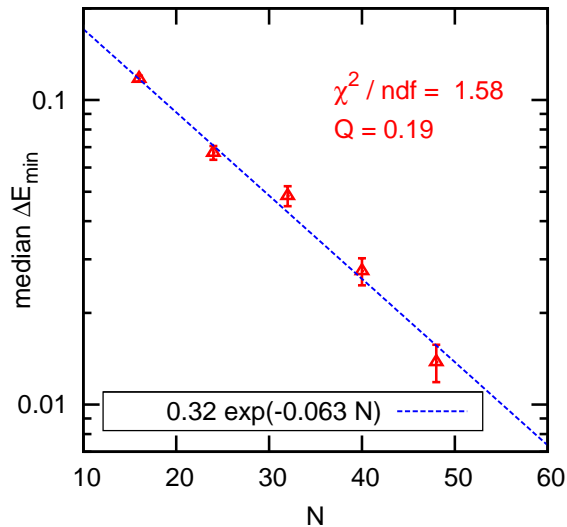}
\caption{(Color online) 
Median gap on a log-linear scale for the locked 2-in-4 problem.
The straight-line fit is good indicating the exponential complexity for this
problem.}
\label{fig:L24DE}\vspace{-0.7cm}
\end{center}
\end{figure}
\begin{figure}
\begin{center}
\includegraphics[width=\figurewidth]{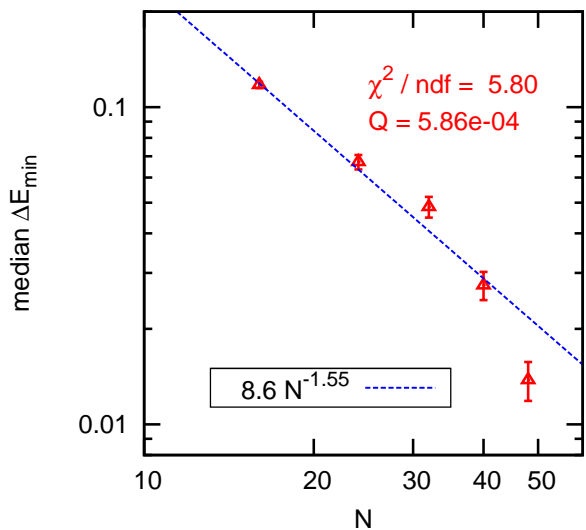}
\caption{(Color online) 
Median gap on a log-log scale for the locked 2-in-4 problem.
The straight-line fit is poor indicating that the complexity for this
problem is not polynomial.}
\label{fig:L24DE_loglog}\vspace{-0.7cm}
\end{center}
\end{figure}

The corresponding results for the locked 2-in-4 problem are shown in
Fig.~\ref{fig:L24DE} (log-lin) and
Fig.~\ref{fig:L24DE_loglog} (log-log). A straight line fit works very well for the
log-lin plot ($Q = 0.19$) but poorly for the
log-log plot ($Q = 5.9 \times 10^{-4}$). This
provides strong evidence that the minimum gap is exponentially small in
the system size.

The 3-regular 3-XORSAT problem has been studied by J\"org {\it et
al.}~\cite{jorg:09} who determined the minimum gap for sizes up to $N = 24$ by
diagonalization, and Farhi {\it et al.}~\cite{farhi:11} who extended the range
of sizes up to $N = 40$ by quantum Monte Carlo simulations. The two sets of
results agree and provide compelling evidence for an exponential minimum gap.
Below we will compare the coefficient in the exponent found by J\"org {\it et
al.} and Farhi {\it et al.} with that for the other models studied here and
with results from a {\em classical} algorithm.

\subsection{Comparison with a classical algorithm}

Since the QAA is designed to serve as an efficient tool for solving
hard optimization problems
%as compared to classical algorithms,
it is
interesting to compare its efficiency with that of a classical algorithm. 

In Ref.~\cite{guidetti:11} it was argued that a reasonable classical algorithm
to compare with QAA is the heuristic local search algorithm known as  WalkSAT
\cite{walksat}, which is similar in spirit to simulated annealing in that both
make moves which reduce the ``energy'', but also sometimes make moves which
increase it to avoid being trapped in the nearest local minimum.

As discussed in Sec.~\ref{sec:intro}, Landau-Zener theory states that, for the QAA,
the computation time is proportional to $1/\Delta E_\text{min}^2$
(neglecting $N$ dependence of matrix elements) and since we find that $\Delta
E_\text{min} \sim \e^{-c\, N}$, the complexity can be written as
\begin{equation}
\mathcal{T} \propto \e^{\mu N} \,,
\label{QAA_complex}
\end{equation}
where $\mu=2\, c$.

In the WalkSAT algorithm, the running time is proportional to the number of
``bit flips'' the algorithm makes (for more details, the reader is referred to
Ref~\cite{guidetti:11}).  Writing the median number of flips as
\begin{equation}
\label{Nflip}
N_\text{flips} \propto e^{\mu N} \,,
\end{equation}
we can now compare the exponent coefficients of the QAA versus those of
WalkSAT.  The latter were measured in Ref.~\cite{guidetti:11}.  For the convenience
of the reader, the values of $\mu$ for both the QAA and WalkSAT are summarized
in Table~\ref{tab:mu}.

\begin{table}
\begin{center}
\begin{tabular}{|c|l|l|c|}
\hline\hline
model        & $\mu$(QAA) &$\mu$(WalkSAT)  & Ratio \\
\hline 
locked 1-in-3 SAT   & 0.084(3)\ \text{(this work)}  &
0.0505(5)\ \,\cite{guidetti:11}  & 1.66 \\
locked 2-in-4 SAT   & 0.126(5) \ \text{(this work)} &
0.0858(8)\ \,\cite{guidetti:11}  & 1.47 \\
3-reg 3-XORSAT      & 0.159(2) \cite{farhi:11}      &
0.1198(20)\,\cite{guidetti:11} & 1.32 \\
\hline\hline
\end{tabular}
\end{center}
\caption{
Values of $\mu$, the coefficient of $N$ in the exponential
complexity of the Quantum Adiabatic Algorithm (QAA),
Eq.~\eqref{QAA_complex},
versus that of
the analogous classical algorithm WalkSAT, Eq.~\eqref{Nflip},
and the ratios between them, for the three problems studied here. The data is
taken from the references shown.}
\vspace{-0.5cm}
\label{tab:mu}
\end{table}

As the table indicates, the exponent coefficients obtained with WalkSAT are
somewhat smaller than those of the QAA, suggesting that the latter algorithm,
in the specific way it was implemented in this paper, is slightly less
efficient than its corresponding classical one for these three problems,
although a problem-by-problem comparison shows that the coefficients are 
fairly similar. It is also evident from the table that the
harder the problem is for WalkSAT, the harder it also is for QAA. 

%Interestingly, out of the problems that we study here by quantum and classical
%general-purpose algorithms, the hardest, 3-XORSAT, is the one which has a polynomial time
%algorithm. The fact that  3-XORSAT is very "glassy" (i.e. very hard to solve
%by general purpose algorithms) while being easy to solve by a special algorithm,
%has been discussed recently to illustrate problems in a claimed proof by
%Deolalikar
%that P is not equal to NP, one of the major unsolved problems in mathematics.
%See for example the discussion
%in Refs.~\cite{ricci-tersenghi:11,wiki_deolikar:11}.

\section{Summary and Conclusions}
\label{sec:conclusions}
Using Quantum Monte
Carlo (QMC) simulations, we studied the complexity of the Quantum Adiabatic
Algorithm (QAA) for three constraint satisfaction problems, two of them in the
NP complexity class -- locked 1-in-3 SAT and locked 2-in-4 SAT -- and one in the
P complexity class -- 3-regular 3-XORSAT.  All three problems show
exponential complexity (albeit with somewhat different coefficients),
i.e.~the
computation time required by the QAA to reach the solution of the problem
Hamiltonian with high probability increases exponentially with the system size
$N$. 

We have also compared the QAA complexities against these of an analogous
classical algorithm, WalkSAT and show the results in
Table \ref{tab:mu}. Perhaps not surprisingly we find that
the harder the problem is for WalkSAT, the harder it also is for the QAA.
Moreover, it seems that the coefficients of $N$ in the exponential in the
expression for the complexity of the QAA, Eq.~\eqref{QAA_complex},
are somewhat larger than
those of WalkSAT, Eq.~\eqref{Nflip} (with ratios ranging between about $1.3$ and $1.7$). 

Several interesting questions arise upon examining the results of this study,
and which we believe would be interesting to study in future work.  The first
one has to do with the possibility of avoiding the exponentially small gap by
repeatedly running the algorithm with different random values for the
transverse fields (and clause costs) \cite{farhi:09}.
It would also be interesting to look,
more generally, for
better paths in Hamiltonian space, perhaps by adding additional terms in the
Hamiltonian for intermediate values of $s$, which would increase the minimum gap.
In particular, can we find a clever way to optimize the path
in Hamiltonian space ``on the fly'' during the simulation? 

While the study reported here used instances with a unique
satisfying assignment (USA), in which case the gap to the first excited state has
a minimum which is related to the complexity, it would be
interesting to also consider {\it random} instances to see if those too have
exponential complexity in QAA. However this is numerically more challenging. 

\begin{acknowledgments}
We thank Eddie Farhi, David Gosset and Anders Sandvik for helpful comments and discussions. 
This work is supported in part by the National
Security Agency (NSA) under Army Research Office (ARO) contract number
W911NF-09-1-0391, and in part by the National Science Foundation under Grant
No.~DMR-0906366. We would also like to thank the Hierarchical Systems Research
Foundation for generous provision of computer support.
\end{acknowledgments}

\bibliography{refs,comments}

\end{document}